\documentclass[prl,twocolumn,showpacs,preprintnumbers,amsmath,amssymb]{revtex4}

\usepackage{epsfig}

\begin{document}

\title{
Current-driven domain wall motion in thin ferromagnetic wires
} 

\author{S. E. Barnes and S. Maekawa }
\affiliation{Institute for Materials Research, Tohoku University, 
Sendai 980-8577, Japan}

\preprint{LL8931}

\date{\today} 
\begin{abstract}
{
The coupling between a current and a Bloch wall is examined in the half-metal limit of the double exchange model. The conduction electrons transfer angular momentum to the Bloch wall with 100\% efficiency in the absence of pinning. The wall is displaced without distortion with  velocity proportional to the current. In the presence of a pinning potential either the angular momentum is destroyed by the perpendicular component of the anisotropy field or is converted to coherent magnons.
A moving wall has its velocity reduced by  pinning.   The expression for the velocity agrees surprisingly well with experiment.
}
\end{abstract}

\pacs{ 
 75.60.-d, 73.40.Cg, 73.50.Bk, 75.70.Ði
}

\maketitle

Spintronic devices have great technological promise but represent a challenging problem at both an applied and fundamental level. It has been  shown theoretically\cite{100,5} that the direction of a magnetic domain might be switched  using currents alone. Devices designed to use this principle often consist of multilayers of magnetic and non-magnetic materials. The advantages of similar devices based upon the  current induced displacement of Bloch wall are simplicity and the fact that the switching current is smaller\cite{1,2,3}. Experimentally the current induced displacement of a Bloch wall has been clearly demonstrated and in a recent experiment the velocity of the displaced wall was measured\cite{3}.

In this Letter is  developed a first principles theory for the coupling between a current and a Bloch wall based upon a standard model for ferromagnets, i.e., within the double exchange model. Great care is taken in order to ensure that angular momentum is conserved. Using the $s-d$-exchange model, Berger\cite{5} shows the polarization of the $s$-electrons rotates as it passes through a wall. The reaction force might then be assumed  to cause wall movement. However, as  will be shown, the situation is complicated. Angular momentum given to the wall can be taken up either  by (i) the displacement of the wall,  (ii) the production of coherent spin waves  or  (iii)  a rotation of the wall so that the angular momentum is absorbed by the anisotropy energy. These issues are most easily addressed  in the  half-metal limit. This is technologically pertinent since it corresponds to the most efficient devices. Most work has been performed on Permalloy (Ni$_{81}$Fe$_{19}$). For this material, the current $\vec j$ has a polarization $p\approx 0.7$\cite{3} and so the present description based upon $p=1$ is not hopelessly unrealistic. When making comparison with experiment, the difference will be accounted for by adding factors of $p$ as appropriate. 

The standard double exchange model is
\begin{eqnarray}
{\cal H}
= - t\sum_{<ij>\sigma}\left(c^\dagger_{i\sigma}c^{}_{j\sigma} + H.c.\right) + U\sum_{i}n_{i\uparrow}n_{i\downarrow} 
\nonumber \\
 - \sum_{i} (J_H \vec S_i\cdot \vec s_i +A {S_{iz}}^2 - K_\perp {S_{iy}}^2)
- J_0 \sum_{<ij>}
\vec S_i\cdot \vec S_j 
\label{un}
\end{eqnarray}
where $\vec S_i$ is the localized spin operator and 
$c^\dagger_{i\sigma}$ creates  an electron  with spin $\sigma$ at site $i$. The wire is assumed to extend along the $z$-direction. That this an easy axis  implies $A>0$. 

Corresponding to  highly correlated electrons, the limit $U \gg t$ is taken.  This permits a spinless majority electron operator $c^\dagger_{i} \equiv c^\dagger_{i\uparrow}$ to be defined, and in terms of this the minority 
$c^\dagger_{i\downarrow} = s^-_i c^\dagger_{i}$ where $s^-_i$ is the conduction electron spin operator.   A large Hund's rule coupling $ |J_H| \gg t$ is also assumed. With this only the largest total angular momentum manifold  is relevant. At a site occupied by a conduction electron only states with $J=S+1/2$ need to be accounted for and it follows that $\vec s_i \approx \vec S_i/(2S)$. This is {\it not\/} an assumption about local equilibrium of the conduction electron but rather is an immediate and {\it rigorous\/} consequence of the Wigner-Eckart theorem.

The Holstein-Primakoff transformation\cite{6} $S_{iz}=S-b^\dagger_i b^{}_i$, $S^+_i = (2S- b^\dagger_i b^{}_i)^{1/2}b^{}_i \approx (2S)^{1/2}b^{}_i$ is used to quantize the spin degrees of freedom. In principle, the axes of quantization are arbitrary, {\it however\/}, as the standard theory of ferromagnetic magnons illustrates, {\it in fact\/} the local axis of quantization must be taken along the {\it classical\/} equilibrium direction, i.e., the spin direction in the limit $S\to \infty$. When this is the case the approach generates and expansion in $1/S$. This  local equilibrium direction is specified by the Euler angles $\theta_i$ and $\phi_i$. In this Letter the solution will most often be time dependent. Rather than setting up the complicated equations of motion, here extensive use will be made of {\it rotating frames}. The aim is to reduce the problem to that of {\it static\/} equilibrium but within the time dependent frame. The time dependent solution is then obtained upon reverting to the laboratory frame.

In the absence of a wall, for large $U$, the kinetic energy term is $\hat T = -t\sum_{\langle ij\rangle} (c^\dagger_i c^{}_j+ s^-_i c^\dagger_i c^{}_j s^+_j)+H.c.$ The Bloch wall is generated through via the {\it static\/} rotations by $\theta$, and $\phi$. For this large $U$ regime, the usual SU(2) rotation of the conduction electron operators simplifies to $c^\dagger_i \to (\cos [\theta/2] + \sin [\theta /2]\,s^-_i)  c^\dagger_i$ and the rotated $\hat T \approx -t\sum_{\langle ij\rangle}\cos [(\theta_i- \theta_j)/2] (c^\dagger_i c^{}_j+ s^-_i c^\dagger_i c^{}_j s^+_j)  - i t\sum_{\langle ij\rangle}\sin [(\theta_i- \theta_j)/2]( s^-_i c^\dagger_i c^{}_j - c^\dagger_i c^{}_j s^+_j)+H.c.$ Charge motion arises principally via ${\cal H}_t  = -t\sum_{\langle ij\rangle}t_{ij} c^\dagger_i c^{}_j +H.c.$ where $t_{ij}=t\cos [(\theta_i- \theta_j)/2]$. The reduction of $t_{ij}$ in the Bloch wall implies a barrier, however this has a height $\sim SA$ which is taken to be negligible compared to $E_F$. The solutions of ${\cal H}_t$ are, to a good approximation, plain $\vec k$-states independent of the wall position or its motion.

Using the Wigner-Eckart theorem, the part of  $\hat T$ linear in $s^\pm_i$ reduces to:
\begin{equation}
{\cal H}_{tJ} = 
-\frac{t}{2(2S)^{1/2}}\frac{\partial \theta_i }{\partial z}a 
\sum_i c^\dagger_i c^{}_{i+1} (b^{}_i - b^{\dagger}_{i+1}) + H.c.,
\label{deux}
\end{equation}
which is the key spin-charge current interaction. The, bi-linear in $s^\pm_i$, part of $\hat T$ is also of importance. When acting on eigenstates of ${\cal H}_t$ these lead to a renormalization, $J = J_0 + (x^\prime t/2S^2)$, of the total exchange coupling $J$. The effective concentration $x^\prime = (1/N)\sum_{\vec k}(\cos k_z a)\, n_{\vec k}$, assuming isotopic hopping.

A small wall {\it translation\/} by $\Delta z$ amounts to a small rotation generated by $R(\Delta z)=\prod_i \exp(i(\theta_i - \theta_{i+1})(\Delta z/a)S_{iy})\approx  (1+ (1/2)\sum_i (\partial \theta_i /\partial z)\Delta z(2S)^{1/2}(b^\dagger_i - b^{}_i))\equiv 1+S^{1/2}(2A/J)^{1/4}(\Delta z/a)(b^\dagger_w-b^{}_w)$ which defines $b^\dagger_w = (J/8A)^{1/4}\sum_i (\partial \theta_i /\partial z)a b^\dagger_i$. 
That the translation $R(\Delta z)$ leaves the energy unchanged implies that $b^\dagger_w$
creates a {\it zero energy\/} excitation localized in the wall. This special $b^\dagger_w$-mode is key to a understanding of not only the dynamics but also pinned static solutions of a Bloch wall. It is of the nature of a discrete Goldstone boson associated with the translational invariance of the wall. 

The effective, ${\cal H} \approx E\{\theta_i, \phi_i\} + {\cal H}_t + {\cal H}_s + {\cal H}_{tJ}$, where $E\{\theta_i, \phi_i\}$ is the classical energy functional, ${\cal H}_t$ is given above and ${\cal H}_s$ is the magnon Hamiltonian. The Bloch wall structure can be determined by minimizing $E\{\theta_i, \phi_i\}$. The standard result is $\theta(z) =2\cot^{-1} e^{-(z/w)}$ with $\phi_i =0$ for $K_\perp >0$. The wall width $ w= a\left(\frac{J }{ 2A}\right)^{1/2}$. The local axes  change from $\theta = 0$ to $\theta = \pi$ with increasing $z$ and the wall lies in the $x-z$-plane.

The conservation of the spin angular momentum $J_z$ is a central issue. If the coupling to the contacts is ignored and $K_\perp =0$, it follows that $[J_z, {\cal H}]=0$. A wall with speed $v$ implies $\dot J_z = \hbar S v /a$ and, {\it in fact\/} this must be equal to the surface integral $\int  \vec s \cdot d \vec A$ where $\vec s $ is the {\it spin\/} current. The conduction electrons have their spin reversed in the wall implying a contribution $\int  \vec s \cdot d \vec A \propto I$, the {\it charge\/} current. Such a finite surface term gives rise to a {\it local\/} term $\nabla\cdot  \vec s \propto j$, the {\it charge current density}, in the equation for $d \vec S_i/dt$ and will lead to wall motion even when $K_\perp =0$.

Whatever the value of $K_\perp$, the translational invariance of the Hamiltonian implies that the wall will move with a constant velocity in the presence of a charge current. In order to prove this, use is made of a rotating frame. Consider again the current carrying states which are eigenstates of $\hat n_{\vec k} = c^\dagger_{\vec k}c^{}_{\vec k}$. The spin-charge coupling reduces to  
\begin{equation}
{\cal H}_{tJ} = 
 -i \frac{ \hbar j a^2 }{ 2 e S}
  \sum_i \frac{ \partial \theta_i }{\partial z}  a  (2S)^{1/2}  (b^{}_i - b^{\dagger}_{i}).
\label{trois}
\end{equation}
Here $(2S)^{1/2}(1/2i)  (b^{}_i - b^{\dagger}_{i}) \approx  S^\ell_{iy}$ defined to be strictly perpendicular to the {\it instantaneous\/} axis of quantization. The only derivative which is finite during this short period is the similar instantaneous $\dot S^\ell_{ix} = S \frac{ \hbar j a^2 }{ 2 e S}\sum_i \frac{ \partial \theta_i }{\partial z}  a$. This derivative can also be made to be zero by a suitable rotating frame, i.e., via $r= \exp (i(\partial \theta_i /\partial t)S^\ell_{iy} t)$ so that $S^\ell_{ix} \to r S^\ell_{ix}r^{-1}$ and $\dot S^\ell_{ix} \to \dot S^\ell_{ix} + i(\partial \theta_i /\partial t) [S^\ell_{iy},S^\ell_{ix}] = \dot S_{ix} - i(\partial \theta_i /\partial t)S$. The wall is therefore stationary,  when the local axes of quantization rotate according to
\begin{equation}
\frac{\partial \theta_i }{ \partial t} = v_0 \frac{\partial \theta_i }{ \partial z};
\ \ \ \ v_0 = \frac{j a^3 }{2 eS}.
\label{quatre}
\end{equation}
The solution of this is $\theta_i \equiv \theta_i(z-v_0t)$ where $\theta_i(z)=2\cot^{-1} e^{-(z/w)}$ is the static solution. In the laboratory frame, the wall is translated {\it without distortion}. The velocity $v_0$ \cite{3,7} is such that the net conduction electron spin current which enters the wall is exactly compensated by the similar current implied by its uniform displacement.

The simplest model for the {\it pinning\/} of the Bloch wall is to reduce by a small amount the anisotropy energy at the origin. The added  term ${\cal H}_p = +\delta A {S_{0z}}^2$  generates a {\it pinning gap\/} $E_p$. The problem of such a pinned wall is surprisingly complicated. Our results developed below show, for finite $K_{\perp}$, the wall tilts by an angle $\phi$ and the critical charge current $j_c$ is determined by the condition that $E_p=0$.  However,  when $K_{\perp}$ is negligible, the wall generates magnons which take away the conduction electron  angular momentum given to the wall. In this case,  $j_c$ is found to reflect the equality of the rate of creation of {\it linear\/} magnon momentum to the maximum force generated by the pinning.

The largest pinning gap $E_p$ arises when the wall is centered at the origin, i.e., with $\theta_0 = \pi/2$.  Explicitly then, $S_{0z} \to S_{0x}$ and the diagonal part of ${\cal H}_p  = 2S \delta A b^\dagger_{0} b^{}_0$. The value of $E_p$ is determined by evaluating the commutator $[[{\cal H}_p, b^{}_w],b^\dagger_w]$,  which is equivalent to taking the expectation value with respect to the $b^\dagger_w$-magnon wave function, and, with $j=0$, gives,  
\begin{equation}
E_p = S\left(\frac{2A }{ J}\right)^{1/2}  \delta A.
\label{cinq}
\end{equation}
The wall can now {\it tilt\/} so that $\phi$ is finite. The anisotropy $K_{\perp} \sum_i {S_{iy}}^2$ contains $- 3SK_{\perp}\sin^2 \theta_i \sin^2 \phi_i b^\dagger_i  b^{}_i$ which gives the $b_w$-mode an energy shift of $-2SK_{\perp} \sin^2\phi$. The {\it maximum\/} allowed value of $\phi$ therefore corresponds to $E_p = S\left({2A / J}\right)^{1/2}  \delta A -2SK_{\perp} \sin^2\phi =0$. This is related to the current through the terms in $K_{\perp} \sum_i {S_{iy}}^2$ which are linear in $b^{}_i$ and $b^{\dagger}_i$. These are imaginary  and can be written as $- i(J/2A)^{1/4}K_{\perp}(\sin 2\phi )(b^\dagger_w - b_w)$. This cancels ${\cal H}_{tJ}$ if
$
(\hbar j a^2 /  2 e S) = (S/2) \left({J / 2A}\right)^{1/2} K_\perp \sin  2\phi
$
whence eliminating $\phi$, assuming this is small, gives a critical current
\begin{equation}
j_{c1} =   \frac{2eS^2}{ \hbar a^2} \left( \frac{J}{ 2A }\right)^{1/4} \sqrt{ \frac{K_\perp \delta A}{ 2}}.
\label{six}
\end{equation}

If $K_\perp$ is small there is an intrinsic maximum current, 
$
j_r =(eS^2/ \hbar a^2) \left({J /2A}\right)^{1/2} K_\perp,
$
for $\phi = \pi/4$, which will be reached before $E_p =0$. It follows for 
large $\delta A$ that $ j_r< j_{c1}$ and  $K_\perp$ is negligible. In order to conserve angular momentum, the wall must now produce magnons, i.e., be  {\it time dependent}. Sought again are current carrying eigenstates. To the $j=0$ wall  is added a twist characterized by $k(z) \equiv \partial \phi/\partial z$ and the solution is rendered stationary by a rotation of angular frequency $\omega$ about the new axes $\theta_i$ and $\phi_i$. The {\it final\/}  axes of quantization then correspond to $\Theta_i = \theta_i + \beta_i$ and $\phi_i$. The effect of such a rotation is to add an effective magnetic field $ \hbar \omega/g \mu_B $  which has an component $\sin \beta_i \hbar \omega/g \mu_B $ {\it perpendicular\/} to the  $\Theta_i $, $\phi_i$ axis. The twist of the bond $i \to i+1$ by $(\partial \phi_i /\partial z)a$ is introduced by $R_{i,i+1} \equiv e^{i(\partial \phi_i /\partial z)a\sum_{i^\prime \ge i+1}S_{i^\prime z}}$. Equilibrium, in the rotating frame, implies zero coefficients of {\it both\/} $b^\dagger_i$ {\it and\/} $b^{}_i $. The easiest fashion by which to establish these equilibrium conditions is to observe the local $S^\ell_{ix} = (2S)^{1/2}(1/2)(b^{}_i + b^{\dagger}_i )$ while $S^\ell_{iy} = (2S)^{1/2}(1/2i)(b^{}_i - b^{\dagger}_i )$. Given that, e.g., $R_y = e^{i S^\ell_{iy}\delta \theta}$ rotates the $i$th spin by a small $\delta \theta$ it is easy to show by evaluating $[S^{\ell}_{iy}, {\cal H}]$ that the coefficient of $S^\ell_{ix}$ in $\cal H$ must be $(1/ 2S)(\partial E /\partial \theta_i)$ where $E\{\theta_i, \phi_i\}$ is the already defined  classical energy. The exchange part of this is $E_J  =  -J\sum_i 
[\cos \Theta_i \cos \Theta_{i+1} + \sin \Theta_i \sin \Theta_{i+1} (\cos \phi_i \cos \phi_{i+1} + \sin \phi_i \sin \phi_{i+1} )]$. A small rotation by $\delta \phi$ is generated rather by $R_z = e^{i S^{}_{iz}\delta \phi}$ where $S^{}_{iz} = \cos \theta_i S^{\ell}_{iz} + \sin \theta_i S^\ell_{ix}$. However $[S^{\ell}_{iz}, {\cal H}]$ contains no $c$-numbers and so this rotation determines the coefficient of $S^\ell_{iy}$, i.e., this is $(1/ 2S\sin \theta_i)(\partial E /\partial \phi_i)$. That the coefficient of $S^\ell_{ix}$ be zero implies 
$
[A S\sin 2\Theta_i  + JS a^2  (\partial^2 \Theta_i/\partial z^2) 
+ (JS a^2)  (\partial \phi_i/\partial z)^2 \sin 2\Theta_i
 + \hbar \omega\sin \beta_i ] \break = 0 
$. This can  be re-integrated by $\Theta_i$ to give back the energy to within a constant, i.e., it is implied that: 
\begin{eqnarray}
[(A S   + \frac{JS a^2}{2} k(z)^2 )\sin^2 \Theta_i &+& (1/2)JS a^2  (\partial \Theta_i/\partial z)^2
\nonumber \\
 - \hbar \omega (\cos \beta_i -1)] &=& 0.
\label{sept}
\end{eqnarray} 
Using  $\partial E_J/\partial \phi$ for the coefficient of $S^\ell_{iy}$ is obtained
$$
\frac{ \partial \Theta_i}{\partial z}
+ \frac{JS a^2}{2} 
\left[ (\sin \Theta_i)^{-1} 
\frac{\partial }{\partial z} k(z) \sin^2\Theta_i 
\right]=0.
$$
Integrating this with respect to $z$ results in an equation relating currents:
\begin{equation}
(JS a) k(z) \sin^2 \Theta_i+ \frac{\hbar j  a^2 }{eS}\cos \Theta_i 
= \frac{\hbar j  a^2 }{eS}
\label{huit}
\end{equation}
and gives
\begin{equation}
k(z) = \frac{1}{2(JS a)\cos^2 \frac{\Theta_i}{2}} \frac{\hbar j  a^2 }{eS}
\label{neuf}
\end{equation}
which, with Eqn.~(\ref{sept}), gives an equation for $\Theta_i$.  However, except to the extreme right of the wall, $\Theta_i \approx \theta_i$ and Eqn.~(\ref{sept})  gives directly $\beta_i$ while Eqn.~(\ref{neuf}) determines the twist $k(z)$. For the extreme right Eqn.~(\ref{neuf}) reduces to $(JS a) k_s {\beta_i }^2 = ({\hbar j  a^2 /eS})$ which equates spin currents and ultimately determines amplitude $\beta_i$ of the magnons leaving the system to the right, while Eqn.~(\ref{sept}) reduces to $\hbar \omega = 2AS + JS a^2 {k_s}^2$ which, correctly, equates the frequency of the rotating frame to the magnon energy.  The whole solution is then parameterized by the value of $k_s$. This, in turn, is determined by requiring that the absolute value of the ground state energy be a minimum. Given that the system is large compared to the width of the wall, the energy to be minimized it that of the coherent magnons to the right of the wall. The result is $k_s a = (2A/J)^{1/2}$ which corresponds a magnon energy per unit site of $e_r =   (2A/J)^{1/2} (\hbar j  a^2 /e)$.

A relationship between $e_r$ and the force on the  wall is obtained by considering a  wall displacement $\Delta z$  generated by $R(\Delta z)=1+S^{1/2}(2A/J)^{1/4}(\Delta z/a)(b^\dagger_w-b^{}_w)$. For the {\it displaced\/} wall ${\cal H}R|0\rangle = [E_0 + e_r (\Delta z/a)]R|0\rangle$, {\it not\/} accounting for pinning. It follows that 
$\Delta E =  [{\cal H},R(\Delta z)] =  e_r  (\Delta z/a)$ and that there is force $f = \Delta E/\Delta z = e_r/a$. Using $[{\cal H},R(\Delta z)] = 
S^{1/2}(2A/J)^{1/4}(\Delta z/a) [{\cal H},(b^\dagger_w-b^{}_w)] =  e_r  (\Delta z/a)$ shows  ${\cal H}$ contains $ (2S)^{-1/2} (2A/J)^{1/4} (\hbar j  a^2 /e) (b^\dagger_w + b^{}_w)$. A similar term arises from the linear part $\tilde P = S\delta A\sin \theta_0 \cos \theta_0 (2S)^{1/2} (b^\dagger_0 + b^{}_0)$ of the pinning term, i.e., $[\tilde P, b^{}_w] =
S^{3/2}(J/2A)^{1/4}S\delta A\sin \theta_0 \cos \theta_0 (\partial \theta_0/\partial z)a $. Corresponding to $\theta_0 = 55^o$, the maximum pinning force, is $(4/3\sqrt{3}) S^{3/2} \delta A (b^\dagger_w + b^{}_w)$. Finally, the magnon de-pinning current is obtained:
\begin{equation}
j_{c2} = \frac{4}{3\sqrt{3}} \left(\frac{ e S^{2}}{\hbar a^2}\right){\delta  A}.
\label{dix}
\end{equation}
This key result is equivalent to equating the rate of creation of linear  magnon momentum to the maximum force produced by the pinning potential.

The voltage across the wall is easily deduced. For large $S$, a non-polarized contact will inject majority $\uparrow$ electrons on the left and this defines $\mu_\uparrow=\mu_\ell$ as the chemical potential of the left contact. For an eigenstate, the energy to add the last $\uparrow$ electron cannot depend on position and thus $\mu_\uparrow$ is {\it the\/} chemical potential for these electrons. One could imagine measuring $\mu_\uparrow$ on the {\it right\/} with a spin polarized contact. The destruction of such an electron involves destroying a magnon and hence $\mu_\uparrow$ must be exactly $\hbar \omega_{\vec k}$ higher in energy than $\mu_\downarrow = \mu_r$,  i.e., the majority and contact chemical potential to the right, i.e, $\mu_\ell - \mu_r = eV = \hbar \omega_{\vec k}\approx 4AS$ for small currents.  There is evidently a branch imbalance.

By design, Permalloy has a small intrinsic anisotropy and both $A$ and $K_\perp$ might be expected $\sim 2\pi M_0$ typically $\sim 0.1$T. This is several orders of magnitude smaller than $J$ and it must be expected that, e.g., variations in the wire width lead to variations $\delta J_i$ and pinning. The largest effect occurs for the components of $\delta J_i$ with a length scale which are comparable to $w$ the wall width. In this case the pinning gap $E_p$ can be as large as $\sim S\sqrt{AJ}$. With this it would be the case that $ j_r < j_{c1} $ and the time dependent solution with $j_{c2}$ will apply.

\begin{figure}[t!]
	\vglue -0.4in
\centerline{\epsfig{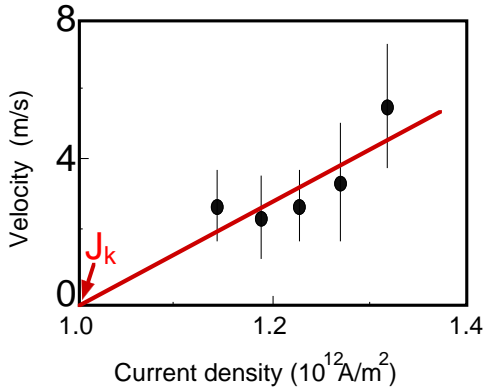}  } 
\vspace{-3pt} 
\caption[toto]{The experimental points are taken from [3]. The solid line is estimated to correspond to the value of $C$  from  Eqn.~(\ref{dixneuf}) and to a $j_k \approx 1.0\times 10^{12}$A/m$^2$. (This has been corrected for the fact that the experimental material has a polarization $p\approx 0.7$.) 
Evidently to within the errors the conduction electron angular momentum, not destroyed by the pinning center is, transmitted to wall motion with efficiency equal to $p\approx 0.7$.  
}
\vglue -0.2in
\label{F1}
\end{figure}

Having considered free motion and pinning it is easy to solve the problem of {\it slow\/} motion with a pinning potential.  If a wall is moving slowly and there are many pinning centers, the wall will {\it adiabatically\/} adjust to remain in the ground state. The wall width is $w=(J/2A)^{1/2} a$ and therefore the characteristic time the wall, with velocity $v$, interacts with the pinning center is $\Delta t = w/v$ and corresponds to an energy $E_c \sim \hbar /\Delta t \sim \hbar (2A/J)^{1/2}(v/a)$. Adiabacity requires that this be less that the energy of the magnons which are involved in the pinning deformations. At the worst these have a $k \sim 1/w$ and the energy involved is $\sim  2SA$ the gap in spin wave spectrum. Adiabatically then implies  $v < v_c \sim 2S(a/\hbar)(AJ)^{1/2}$ which is $\sim (A/J)^{1/2}$ times the maximum spin wave velocity. Given the experimental values\cite{3} correspond to $v \sim 3$m/s this criterion is surely satisfied. 

The principle of the calculation is to 
divide into two parts the rotating frames used to render the Bloch wall stationary, one of which corresponds to the translational motion and a second which reflects the pinning. The displacement corresponds to the rotation operator $R(\Delta z) = \prod_i (1+ (1/2)(\partial \theta_i /\partial z)\Delta z(2S)^{1/2}(b^\dagger_i - b^{}_i))$ where $\Delta z = {v}  t$. Given the adiabatic approximation is well satisfied the  rotations $R({v}  t)$ can be absorbed into {\it slowly\/} time dependent conduction electron operators $R(vt) c^{}_i {R}^{-1}(vt) $. In this moving frame, the effective current density is {\it reduced\/} to $j^\prime = j(1- \frac{v}{v_0})$. Then following the calculation for pinning,  the remaining rotations $R^p_i$ are those required for the pinned ground state. The resulting $j^\prime $ is given by Eqn.~(\ref{dix}) and is determined by the pinning strength. The difference between the actual current and that which can be pinned is $j-j^\prime$, and this in turn, determines $v$ and its relationship with the current $j$. The velocity $v$ is given by:
\begin{equation}
v = C (j-j_k);
 \ \ \  \  C \equiv \frac{  a^3 }{2 e S}
\label{dixneuf}
\end{equation}
where, as stated, $j_k$ is the $j_c$ given by Eqn.~(\ref{dix}) {\it except\/} that average rather than maximum pinning strength is involved. The important result here is that $C$ is independent of $j_k$, i.e., that part of the angular momentum current which is in excess of that converted to magnons by the pinning potential is 100\% converted into motion of the wall. In Fig.~(\ref{F1}) this prediction, corrected for $p=0.7$, is compared with the experiments of Yamaguchi et al. Experimentally it is observed that there is a minimum velocity of $\sim 3$m/s for which uniform motion is observed.  It is reported that no motion of the wall is seen for currents smaller that $1.0\times 10^{12}$A/m$^2$ and taking this to be $j_k$ implies the solid line shown in the figure with a  gradient which would correspond to $\sim pC$. 
What is strange is that the same velocity is reported for a currents in the range $1.15\times 10^{12}$A/m$^2$ to $1.25\times 10^{12}$A/m$^2$. It is unfortunate that it is not possible to follow the system to higher velocities in order to see if the initial points have to do with threshold effects, e.g., reflecting the first order nature of the transition to the magnon producing pinned state. The wall  will be displaced during the time that the magnon amplitude is developed and, in a pulse experiment, this might be interpreted as depinning.

This work was supported by a Grant-in-Ad for Scientific Research on
Priority Areas from the Ministry of Education, Science, Culture and
Technology of Japan, CREST and NAREGI. The  SEB is on a temporary leave from the
Physics Department, University of Miami, FL, U.S.A. and wishes to
thank the members of IMR for their kind hospitality.

\end{document}